\documentclass[aps,prl,reprint,amsmath,amssymb,superscriptaddress,floatfix]{revtex4-2}

\usepackage[colorlinks, citecolor=blue, linkcolor=blue]{hyperref}
\usepackage{xr}
\usepackage{graphicx}% Include figure files
\usepackage{textcomp}% for textmu
\usepackage{xcolor}% for textmu
\usepackage{import}% for textmu
\usepackage{dcolumn}% Align table columns on decimal pointhttps://www.overleaf.com/project/5e7e1af09f08b8000122d056
\usepackage{bm}% bold math

\usepackage{orcidlink}
\usepackage[utf8]{inputenc}

\begin{document}

\title{Entropy-Assisted Nanosecond Stochastic Operation \\in Perpendicular Superparamagnetic Tunnel Junctions}

\author{Lucile Soumah\orcidlink{0000-0002-8287-7310}} 
\thanks{L. S. and L. D. contributed equally.} 
\affiliation{Univ. Grenoble Alpes, CEA, CNRS, Grenoble INP, SPINTEC, 38000 Grenoble, France}

\author{Louise Desplat\orcidlink{0000-0002-8968-1046}} 
\thanks{L. S. and L. D. contributed equally.}
\email{louise.desplat@cea.fr}
\affiliation{Univ. Grenoble Alpes, CEA, CNRS, Grenoble INP, SPINTEC, 38000 Grenoble, France}
\affiliation{Nanomat/Q-mat/CESAM, Universit\'e de Li\`ege, B-4000 Sart Tilman, Belgium}

\author{Nhat-Tan Phan\orcidlink{0000-0002-2806-1660}} 
\affiliation{Univ. Grenoble Alpes, CEA, CNRS, Grenoble INP, SPINTEC, 38000 Grenoble, France}

\author{Ahmed Sidi El Valli\orcidlink{0000-0003-1106-2664}}
\affiliation{Univ. Grenoble Alpes, CEA, CNRS, Grenoble INP, SPINTEC, 38000 Grenoble, France}

\author{Advait Madhavan\orcidlink{0000-0002-4121-1336}}
\affiliation{Associate, Physical Measurement Laboratory, National Institute of Standards and Technology, Gaithersburg, Maryland 20899, USA} 
\affiliation{Institute for Research in Electronics and Applied Physics, University of Maryland, College Park, Maryland 20742, USA}

\author{Florian Disdier\orcidlink{0000-0003-3513-2241}} 
\affiliation{Univ. Grenoble Alpes, CEA, CNRS, Grenoble INP, SPINTEC, 38000 Grenoble, France}

\author{Stéphane Auffret\orcidlink{0009-0006-5008-227X}} 
\affiliation{Univ. Grenoble Alpes, CEA, CNRS, Grenoble INP, SPINTEC, 38000 Grenoble, France}

\author{Ricardo C. Sousa,\orcidlink{0000-0001-8903-3359}} 
\affiliation{Univ. Grenoble Alpes, CEA, CNRS, Grenoble INP, SPINTEC, 38000 Grenoble, France}

%\author{Mark D. Stiles} 
%\affiliation{Physical Measurement Laboratory, National Institute of Standards and Technology, Gaithersburg, Maryland 20899, USA}

\author{Ursula Ebels,\orcidlink{0000-0001-5061-5538}} 
\affiliation{Univ. Grenoble Alpes, CEA, CNRS, Grenoble INP, SPINTEC, 38000 Grenoble, France}

\author{Philippe Talatchian,\orcidlink{0000-0003-2034-6140}}
\email{philippe.talatchian@cea.fr} 
\affiliation{Univ. Grenoble Alpes, CEA, CNRS, Grenoble INP, SPINTEC, 38000 Grenoble, France}

\begin{abstract}
We demonstrate a good agreement between mean dwell times measured in 50~nm diameter, perpendicularly magnetized superparamagnetic tunnel junctions (SMTJ), and theoretical predictions based on Langer's theory. Due to a large entropic contribution, the theory yields Arrhenius prefactors in the femtosecond range for the measured junctions, in stark contrast to the typically assumed value of 1~ns. Thanks to the low prefactors, and fine-tuning of the perpendicular magnetic anisotropy, we report measured mean dwell times as low as 2.7~ns under an in-plane applied field at negligible bias voltage. 
Under a perpendicular applied field, we predict a Meyer-Neldel compensation phenomenon, whereby the prefactor scales like an exponential of the activation energy, in line with the exponential dependence of the measured dwell time on the field. 
We further predict the occurrence of (sub)nanosecond dwell times as a function of effective anisotropy and junction diameter at zero bias voltage.
These findings pave the way towards the development of ultrafast, low-power, unconventional computing schemes operating by leveraging thermal noise in perpendicular SMTJs, which can be scaled down below 20~nm.

\end{abstract}

\maketitle

%a shorter version because intro is very long: yep OK for me.
In magnetism, mastering thermal activation is essential for understanding the complex interplay of temperature and dynamics that governs the behavior of spintronic nanostructures. While enhancing thermal stability is crucial for preventing information loss in nonvolatile memory~\cite{bhatti2017spintronics, dieny2020opportunities,kent2015new}, reduced stability facilitates rapid state switchings, enabling energy-efficient cognitive computing schemes~\cite{grollier2020neuromorphic,Cai_Review_MTJ_2023}. 

Magnetic tunnel junctions (MTJs) are a prime example of a device in which the study of thermal activation is important. MTJs consist of two ferromagnetic layers separated by an insulating oxide, and exhibit two metastable states, corresponding to the relative orientations of the magnetization in the two layers, namely, parallel (P), or antiparallel (AP). These states, readable through the tunneling magnetoresistance (TMR), have distinct resistance levels, and are separated by an energy barrier. In particular, we refer to junctions in which thermal fluctuations induce random switchings between the two states at a scale of a few seconds and below~\cite{hayakawa2021nanosecond, rippard2011thermal, TRNG_MTJ}, as superparamagnetic tunnel junctions (SMTJs). Despite their inherently stochastic resistance fluctuations, the corresponding state probabilities can be controlled deterministically, either through current-induced spin-transfer torques (STT)~\cite{berger, slonczewski, ralph2008spin}, or via an external magnetic field. This tunability, coupled with their energy-efficiency, has made SMTJs highly appealing for cognitive applications, including stochastic implementations of artificial neural networks~\cite{Daniels2019EnergyefficientSC}, brain-inspired~\cite{mizrahi2018neural}, and probabilistic schemes~\cite{borders2019integer, singh2024cmos}. 

\begin{figure*}[ht!]
    \includegraphics[width=\textwidth]{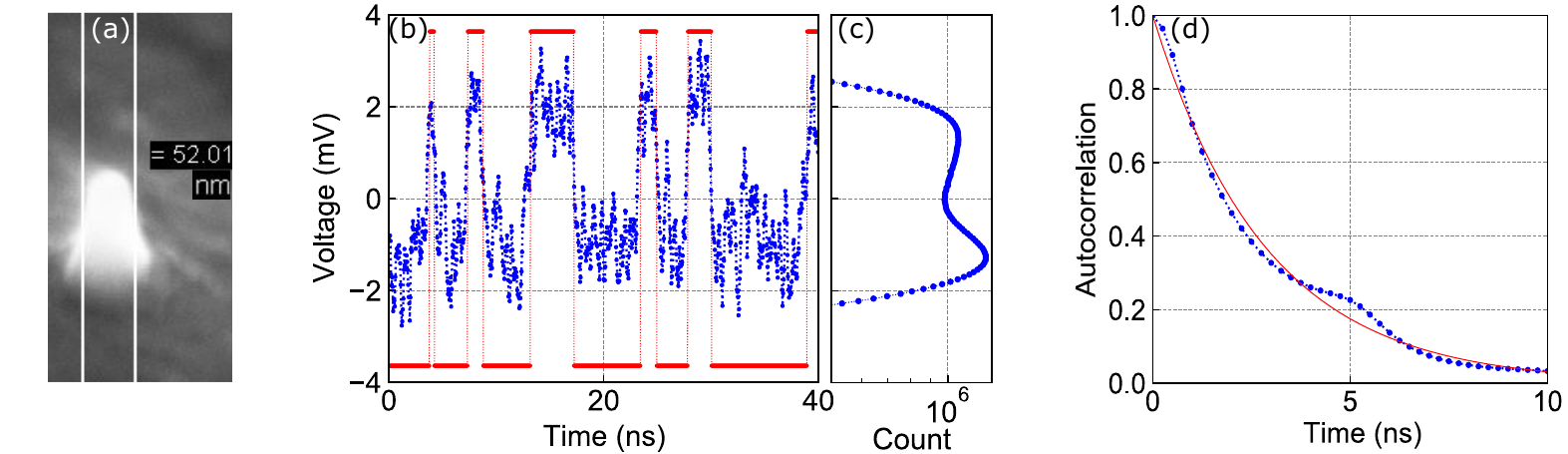}
    \caption{\label{fig_1} (a) Scanning Electron Microscopy (SEM) image of a nominal 50~nm diameter perpendicular SMTJ pillar obtained after ion beam etching. (b) Random telegraph noise signal of a 50~nm diameter perpendicular SMTJ under an in-plane applied field of $\mu_\mathrm{0} $H$_{\parallel}$= 60~mT, and 15~µA mean current (85~mV mean voltage). A 40~ns portion of the recorded voltage-time trace exhibiting 5 transitions is shown in blue, together with its numerically digitized replica, shown in red. (c) Histogram of the overall voltage values measured over 10~ms  and containing more than $10^6$ transitions. (d) Autocorrelation function of the recorded voltage-time trace partially shown in (b) in blue, with an exponential fit shown in red, with 3.6~ns time constant.}
\end{figure*}

A prospective strategy to further reduce energy consumption in these schemes is to lower the dwell times between magnetization reversals,
%, thereby reducing computing time without additional energy or area occupancy overheads.  A straightforward approach is to
by reducing the energy barrier between states to a few~$kT_{\mathrm{RT}}$~\cite{faria2017low}, in which $k$ is Boltzmann's constant, and $T_{\mathrm{RT}}$ is room temperature. So far, typical macrospin-based approaches have suggested that the mean dwell times cannot be lower than an assumed characteristic attempt time, $\tau_0 \approx1$~ns, even with negligible energy barriers, thus implying a notable speed limitation~\cite{kanai2021theory,rippard2011thermal}.
Moreover, claims were made that in-plane MTJs (iMTJs) are inherently faster than perpendicular ones, based on a macrospin model where the large perpendicular uniaxial anisotropy renders the system 2d, implying that only iMTJs can reach nanosecond dwell times~\cite{hayakawa2021nanosecond,kanai2021theory}. This has shifted focus towards iMTJs, which are reported to reach the shortest dwell times of a few nanoseconds, albeit without a complete physical picture~\cite{hayakawa2021nanosecond, safranski2021demonstration, Schnitzspan_stt_joules}. 
However, the thermal stability of iMTJs, dominated by dipolar interactions, makes them sensitive to the junction's shape. When scaled down to nm sizes, the latter becomes hard to control due to patterning-induced defects, resulting in high device-to-device variability.
On the other hand, magnetic properties of perpendicular MTJs are less susceptible to the same shape variability, allowing for downscaling below 20~nm~\cite{cacoilo:hal-02574632, jinnai2020scaling} and integration into foundry industrial processes~\cite{jung2022crossbar, sakhare2018enablement}.

Meanwhile, theoretical studies suggested that the assumption of a constant 1~ns attempt time is not justified. The Arrhenius prefactor of a typical MTJ used for data storage was predicted to vary by many orders of magnitude with material parameters and reach extreme values down to $10^{-21}$~s~\cite{desplat_2020,desplat2020quantifying}, due to a large entropic contribution which is absent in macrospin models. The latter stems from the domain wall mediated reversal pathway, which favors the transition state against the collinear state due to the increased number of available degrees of freedom. This effect was shown to yield a case of the Meyer-Neldel compensation rule, whereby the Arrhenius prefactor scales like an exponential of the activation energy~\cite{peacock1982compensation,yelon1990microscopic,yelon1992origin}. While compensation was predicted to reduce the stability of memory elements, it should be beneficial to achieve faster switching rates for computing.

In this work, we experimentally validate the prediction of extremely low Arrhenius prefactors in magnetic tunnel junctions~\cite{desplat_2020}. By measuring mean dwell times in perpendicular, 50~nm nominal diameter SMTJs under applied in- and out-of-plane magnetic field at low bias voltage, we obtain an excellent agreement with our predictions based on Langer's theory~\cite{langer,coffey2012thermal}. These reveal prefactors at the femtosecond scale, hinting that compensation may be occurring in our measured system. We show that these extremely low prefactor values are instrumental in achieving mean dwell times as short as a few nanoseconds in our experiments. %at increased applied field or bias voltage. 
Finally, we theoretically predict the occurrence of (sub)nanosecond dwell times at zero voltage under a low in-plane field, by reducing the effective perpendicular anisotropy and/or junction diameter.

The stack structure of the fabricated SMTJs consists of Si substrate / SiO$_2$/Ta(15)/Pt(5)/[Co(0.5)/Pt(0.25)]$_6$/
Ru(0.9)/ [Co(0.5)/Pt(0.25)]$_2$/ Co(0.5) / W(0.25) / CoFeB($t_\mathrm{fixed}$) /
MgO(1.25) / CoFeB($t_\mathrm{free}$)/ W(2)/Pt(5).
Numbers in parentheses denote thickness in nanometers, and subscripts on brackets show bilayer repeat counts.  The two uniform, perpendicularly magnetized ferromagnetic layers of CoFeB, i.e., fixed and free, are separated by an insulating MgO layer. In order to reach the superparamagnetic regime while keeping the perpendicular easy axis, we gradually tune the effective magnetic anisotropy as described in the Supplemental Material (SM)~ \cite{sm} through the interfacial anisotropy term $K_{\mathrm{int}}\propto \frac{1}{t_\mathrm{free}}$, by adjusting the free layer thickness $t_\mathrm{free}$ from 1.3 to 1.8~nm.  
Measurements are performed on two different stacks labeled 1 and 2, with nominal free layer thicknesses of $t^{\mathrm{stack 1}}_{\mathrm{free}}=1.21\pm0.20$~nm and $t^{\mathrm{stack 2}}_{\mathrm{free}}=1.16\pm0.20$~nm, and nominal diameter $d=50\pm 20$ nm (see  Fig.~\ref{fig_1}(a)). Correspondingly, the free layer's perpendicular anisotropy can be evaluated for both nanopatterned stacks (see~\cite{sm}).

 We measure voltage fluctuations across the fabricated SMTJs using a high-bandwidth circuit designed to capture submicrosecond events, as described in the SM\cite{sm}. Fig.~\ref{fig_1}(b) depicts voltage-time traces obtained from measurements performed on stack 1 under a 60~mT in-plane magnetic field and 15~µA mean current (85~mV mean voltage). The signal exhibits nanosecond timescale random telegraph noise (RTN). To ensure reliable statistics, we collected more than $10^6$ stochastic transitions, allowing the identification of two distinct voltage levels, as shown in the corresponding voltage histograms in Fig.~\ref{fig_1}(c). These are associated with low and high resistance states, respectively, consistent with P and AP magnetic configurations. The nanosecond scale of the observed mean dwell times is further confirmed in Fig.~\ref{fig_1}(d) by the autocorrelation functions calculated over the entire recorded signal, which exhibit pseudo-exponential decay with time constants of 3.6~ns. %

\begin{figure}
    \includegraphics[width=\columnwidth]{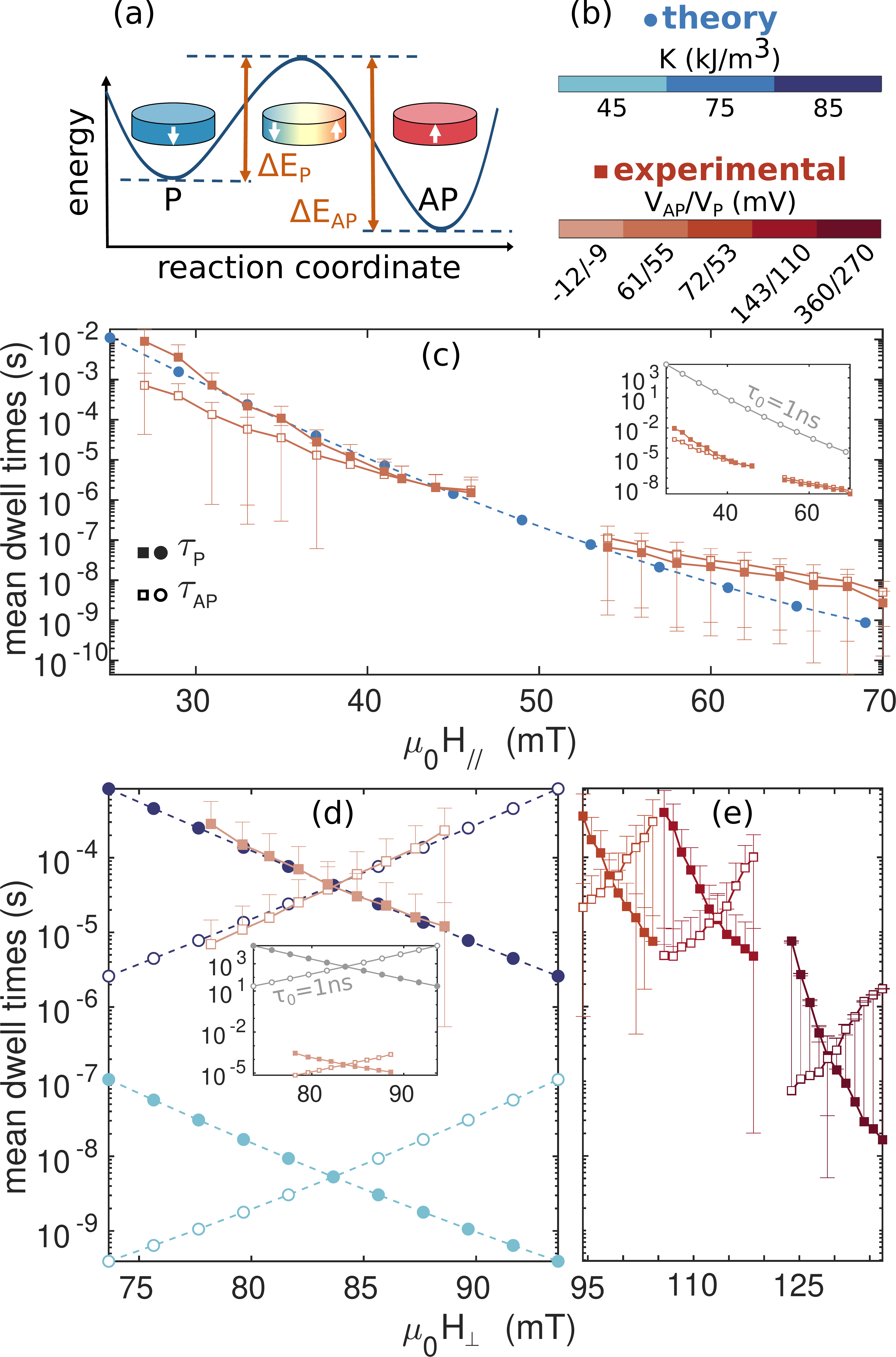}
    \caption{\label{fig_2} (a) Sketch of the energy profile along the reaction coordinate for P and AP states, where the associated energy barriers are indicated as orange arrows. (b)  General legend for (c, d, e). (c, d, e) Experimental (red squares) and theoretical (blue circles) mean dwell times as a function of applied (c) in-plane field for stack 1 (d , e) out-of-plane field for stack 2. The insets in (c, d)
    show a comparison of the measured dwell times with dwell times computed with the barriers in Fig.~\ref{fig:simus}(a, c) and $\tau_0=1$~ns in gray.}    
\end{figure}

In what follows, we demonstrate two different experimental realizations of nanosecond mean dwell times, via in- and out-of-plane applied magnetic fields and currents in perpendicular SMTJs.
We first apply an in-plane magnetic field $H_{\parallel}$ on stack 1, under a minimal %current of 15~µA 
voltage $V_{\mathrm{AP}}/V_{\mathrm{P}}$ of 61/55~mV across the SMTJ--necessary for electrical readout--and vary $\mu_0H_{\parallel}$  from 25 to 70~mT. The field simultaneously reduces the energy barrier, $\Delta E_\mathrm{P,AP}$, for both P and AP states (Fig.~\ref{fig_2}(a)), by reducing the energy of the transition state~\cite{sm}. As shown in red in Fig.~\ref{fig_2}(c), the obtained graph reveals a large reduction of the mean dwell times with the field, from about 10~ms, to nearly 1~ns, without the need for large bias voltage.
Next, we apply a perpendicular magnetic field $H_{\perp}$ on stack 2, which decreases (increases) $\Delta E_\mathrm{P(AP)}$ by increasing (decreasing) the energy of state P (AP). 
To this end, we vary $\mu_0H_{\perp}$ from 75 to 90~mT under a minimal voltage of -12/-9~mV across the MTJ. The resulting mean dwell times are shown in red in Fig.~\ref{fig_2}(d), revealing a limited variation contained between 0.2~ms and 5~µs. To access smaller timescales, we combine the perpendicular field with larger voltages. %1~µA to 40~µA
The results are summarized in Fig.~\ref{fig_2}(e). The obtained graph reveals a vast spectrum of mean dwell times, from about 0.1~ms, to 20~ns at the edge of the magnetic field range, i.e., above 120~mT, for voltage extremes of %40~µA
360/270~mV, which highlights the crucial role of the STT in enhancing the switching rates~\cite{rippard2011thermal, Schnitzspan_stt_joules}. We find the effect of Joule heating on the dwell times to be negligible~\cite{sm}. In terms of power consumption, we  present in the SM a benchmarking of our system against the current state of the art, and show that our SMTJs are highhly competitive, with energy per fluctuation as low as 1.89~fJ~\cite{sm}--a significant reduction compared to the smallest values previously reported in iMTJ~\cite{hayakawa2021nanosecond}.

The dwell times presented in Figs.~\ref{fig_2}(d, e) reveal a seemingly exponential dependence on $H_{\perp}$, in line with the Arrhenius law,
\begin{equation}\label{eq:arrhenius}
\tau_\mathrm{P(AP)} = \tau_0^{\mathrm{P(AP)}} e^{\Delta E_\mathrm{P(AP)}/kT_{\mathrm{RT}}},
\end{equation}
where $\tau_0^{\mathrm{P(AP)}}$ is a prefactor, commonly referred to as an attempt time. For $H_K \gg H_{\perp}$, in which $H_K$ is the anisotropy field, $\Delta E_\mathrm{P(AP)}$ is approximately linear in $H_{\perp}$~\cite{Chaves2015thermal}. Moreover, as we show below, $\tau_0^{\mathrm{P(AP)}}\propto e^{-\Delta E_\mathrm{P(AP)}}$, which yields $\tau_\mathrm{P(AP)}\propto e^{\mp H_{\perp}}$. Note that the barrier dependence on $H_{\parallel}$ is not trivial~\cite{capriata2023energy}, and neither is that of  $\tau_0^{\mathrm{P(AP)}}$, so we cannot conclude on the dwell time dependence on $H_{\parallel}$.

%--------------------------------------------------------------%
%\paragraph*{Simulations} -
To shed light on the mechanism behind the observed dwell times, we perform micromagnetic simulations with a homemade code~\cite{desplat_2020}. For both stacks, we use a spin stiffness $A=10$~pJ/m~\cite{sampaio2016disruptive}, and Dzyaloshinkii-Moriya interaction (DMI) $D=0.14$~mJ/m$^2$~\cite{urrestarazu2024electrical}. For stack 1, we use a saturation magnetization $M_S=1.16$~MA/m, a mesh size equal to the free layer thickness $a=t_{\mathrm{free}}=1.4$~nm, a diameter $d=50$~nm, and effective perpendicular anisotropy $K=75$~kJ/m$^3$. For stack 2, we use $M_S=1.03$~MA/m, $a=t_{\mathrm{free}}=1.1$~nm, $d=40$~nm, and $K=85$~kJ/m$^3$. We also show simulations for $K=45$~kJ/m$^3$. Further discussions on the parameters can be found in the SM~\cite{sm}.
\begin{figure}
\includegraphics[width=1\linewidth]{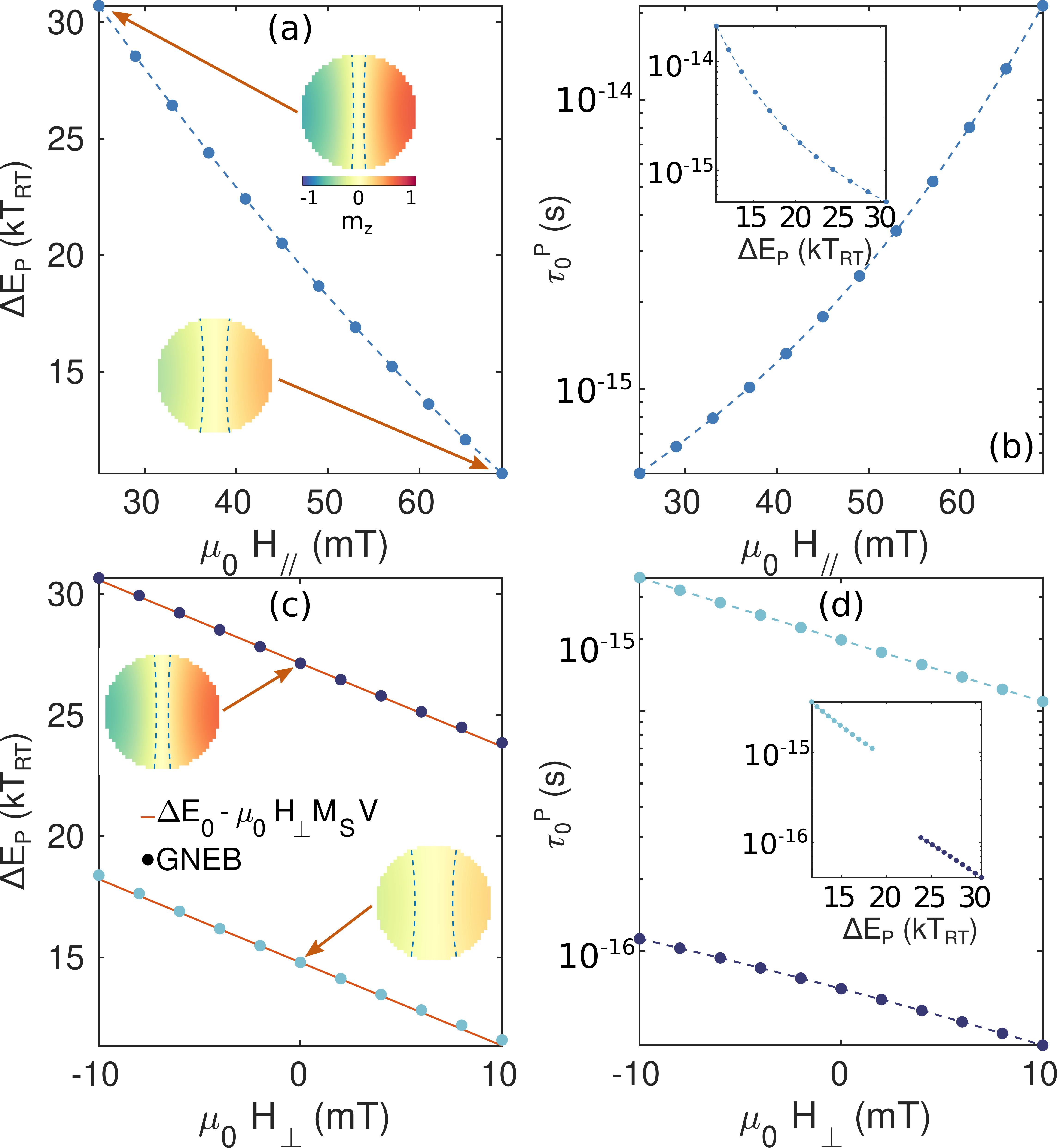}
\caption{Details of the dwell time computations. (a, c) Energy barriers and (b, d) prefactor as a function of (a, b) in-plane field for stack 1 (c, d) out-of-plane field for stack 2. The insets in (a, c) show the spin configuration at the saddle point for different (a) fields and (c) anisotropies, where the dashed isolines correspond to $m_z=\pm 0.1$ to indicate the position of the wall. The insets in (b, d) show the prefactor as a function of the barrier in logarithmic scale, highlighting in (d) the compensation effect. %forgot mz colorbar+add linear fit in Hperp
}
\label{fig:simus}
\end{figure} 

Following previous work~\cite{desplat_2020}, we use the geodesic nudged elastic band method (GNEB)~\cite{bessarab2015method} to compute reversal pathways, which we show in the SM~\cite{sm}, and corresponding energy barriers. The results are gathered in Figs.~\ref{fig:simus}(a, c), respectively for stacks 1 and 2 as a function of in- and out-of-plane field, in units of $kT_{\mathrm{RT}}$ with $T_{\mathrm{RT}}=294$~K. In agreement with previous studies, the reversal occurs through the nucleation and propagation of a domain wall through the disk~\cite{khvalkovskiy2013basic,jang2015detrimental, sampaio2016disruptive,desplat_2020,desplat2020quantifying}. The saddle point configurations with the domain wall in the center are shown as insets for 2 values of $H_{\parallel}$  (Fig.~\ref{fig:simus}(a)), and $K$ (Fig.~\ref{fig:simus}(c)). As shown,  the barriers in Fig.~\ref{fig:simus}(c) scale like $\Delta E_0-\mu_0H_{\perp}M_SV$, where $\Delta E_0\sim \sqrt{AK}dt_{\mathrm{free}}$ is the barrier at 0 field~\cite{Chaves2015thermal}, and $V$ is the volume of the free layer.

The prefactor is computed with Langer's theory~\cite{langer,coffey2012thermal}, as
\begin{equation}
    \tau_0^{\mathrm{P}}
    =\left(\frac{\lambda_+^\mathrm{P}}{2\pi}\Omega_0^\mathrm{P}\right)^{-1},
\end{equation}
in which $\Omega_0^\mathrm{P}$ is the entropic contribution that contains the curvatures of the energy landscape at the initial state and the saddle point, and $\lambda_+^\mathrm{P}$ is the characteristic time associated with the unstable translation of the wall at the saddle point (see SM~\cite{sm} for more details). 

The computed values of $\tau_0^\mathrm{P}$ are presented in Figs.~\ref{fig:simus}(b, d), and are divided by 2 to account for the 2 equivalent saddle points~\cite{desplat_2020}. We find $\tau_{0}^{\mathrm{P}}$ in the 0.1-10~fs range, i.e., up to 7 orders of magnitude lower than the value of 1~ns typically considered in the literature~\cite{weller1999thermal, chen2010advances, lederman1994measurement,Chaves2015thermal,sampaio2016disruptive,kanai2021theory}.
Such values stem from the larger number of states available to thermal fluctuations at the saddle point in the presence of a domain wall, compared to the uniform state, which results in a large entropic contribution. In that sense, the Arrhenius prefactor is not a physical attempt time. 
Moreover, for families of transitions obtained by varying the anisotropy, and the DMI, this system was shown to exhibit Meyer-Neldel compensation~\cite{desplat_2020, desplat2020quantifying}, whereby $\tau_0\propto e^{-\Delta E}$~\cite{peacock1982compensation,yelon1990microscopic,yelon1992origin}.
In the insets of Figs.~\ref{fig:simus}(b, d), $\tau_0^\mathrm{P}$ is shown in logarithmic scale as a function of the energy barriers obtained as a function of $H_\parallel$, and $H_\perp$. For $H_\perp$, the graphs are approximately straight lines, implying that compensation also tends to occur under a perpendicular magnetic field. Along with the linear barrier dependence on  $H_\perp$, this agrees with the observation that the measured dwell times in Figs.~\ref{fig_2}(d, e) scale like $e^{\mp \mu_0 H_\perp M_S V}$. We note that this was also reported in Ref.~\cite{hayakawa2021nanosecond}, but was attributed to a macrospin-like behavior. Under $H_\parallel$, the system does not seem to exhibit compensation, although the prefactor assumes similarly low values. %

The mean dwell times are computed with Eq.~(\ref{eq:arrhenius}), and gathered in Figs.~\ref{fig_2}(c, d, e) along with the experimental values, as a function of in- (Fig.~\ref{fig_2}(c)) and out-of-plane field (Fig.~\ref{fig_2}(d)). Under $H_{\parallel}$, the barrier remains symmetric, so $\tau_{\mathrm{P}}=\tau_{\mathrm{AP}}$. For $H_{\perp}$, we compute $\tau_{\mathrm{P}}$, and note that $\tau_{\mathrm{AP}}(-H_{\perp})=\tau_{\mathrm{P}}(H_{\perp})$. 
In both cases, the computed dwell times agree well with the experimental values obtained at negligible bias voltage, with barriers of about 5 and $25kT_{\mathrm{RT}}$ respectively yielding ns, and $\mu$s dwell times due to the sub-fs prefactors. Additionally, we show that with a lower anisotropy of $45$~kJ/m$^3$, (sub)nanosecond dwell times can also be obtained under a perpendicular field at 0 current.
For comparison, in the insets of Figs.~\ref{fig_2}(c, d), we show that dwell times computed from the GNEB barriers and $\tau_0^{\mathrm{P(AP)}}=1$~ns (in gray) fail to reproduce the measured dwell times (in red) by over 6 orders of magnitude. This discrepancy is even more drastic when using macrospin energy barriers.

\begin{figure}
\includegraphics[width=1\linewidth]{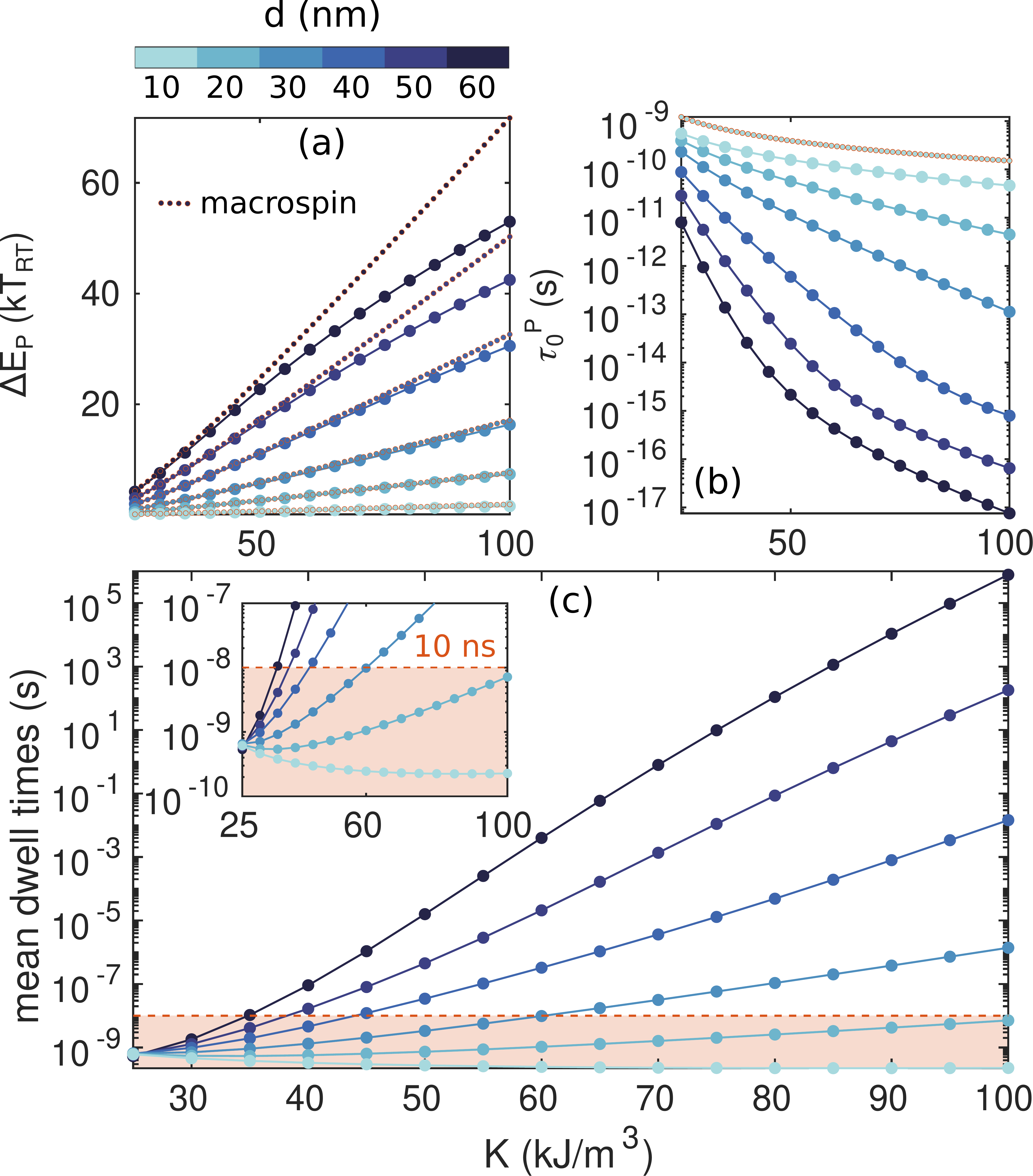}
\caption{Theoretical predictions of (sub)nanosecond dwell times at zero current and 25~mT in-plane field, in a 1.4~nm thick MTJ using the parameters of stack 1. (a) Energy barriers, (b) Arrhenius prefactors, and (c) mean dwell times as a function of effective anisotropy, for junction diameters between 10 and 60~nm. The dotted lines in (a, b) correspond to the macrospin model. The inset in (c) shows a zoom of the dwell times around 1~ns.}
\label{fig_4}
\end{figure}

Last, given the scalability appeal of perpendicular MTJs, and the drive for low energy operation, we compute dwell times for junction diameters from 60 to 10~nm to demonstrate that (sub)nanosecond dwell times can also be obtained at low field and zero currents. We use the parameters of stack 1 and vary the effective anisotropy from 100 to 25~kJ/m$^3$. The in-plane field is set to $25$~mT, so that $m_z\leq 0.8$ for all anisotropy values~\cite{sm}, to ensure electrical detectability. The results are gathered in Fig.~\ref{fig_4}.
The energy barriers and the prefactors are respectively given in Figs.~\ref{fig_4}(a) and (b). The macrospin model is also shown in dotted lines~\cite{sm}, where the prefactor was derived analytically from Langer's theory~\cite{schratzberger2010validation}. Although the barriers coincide with the macrospin picture for $d\lesssim 40$~nm, the prefactor still differs from it by orders of magnitude down to $20$~nm diameter, due to the presence of nonuniform magnon modes. The mean dwell times thus reach the (sub)nanosecond regime for $K\le$35~kJ/m$^3$ for $d=60$~nm, to $K\le60$~kJ/m$^3$ for $d=$30~nm, and up to $K\ge100$~kJ/m$^3$ for $d\le20$~nm.

%%%%%%%%%%%%%%%%%%%%%%%%%%%%%%%%%%

In this work, we demonstrated a good agreement between mean dwell times measured in perpendicular SMTJs with 50 nm nominal diameter at negligible bias voltage, and theoretical predictions based on Langer's theory. The later yields Arrhenius prefactors in the femtosecond range for the measured junctions, due to a large entropic contribution~\cite{desplat_2020}. Under a perpendicular field, we found a case of the Meyer-Neldel compensation, where the prefactor scales like an exponential of the energy barrier. This is experimentally verified by the exponential dependence of the dwell times on $H_{\perp}$.

Our fine-tuning of the effective anisotropy in our junctions, combined with the small values assumed by the prefactor, allowed us to experimentally measure nanosecond mean dwell times as low as 2.7~ns under a 70~mT in-plane field at negligible bias voltage--approximately 4 orders of magnitude shorter than previously reported in perpendicular SMTJs~\cite{bapna2017current,parks2018, funatsu2022local,kobayashi_sigmoid_psmtj}, and similar to values so far reported in in-plane junctions~\cite{hayakawa2021nanosecond,safranski2021demonstration,Schnitzspan_stt_joules}, with even lower energy per fluctuation~\cite{sm}. We also reached comparable values under an effective perpendicular field of $\pm10$~mT at larger voltage, mainly attainable through the effect of STT.

By comparing our dwell time predictions to that of a macrospin model, we further debunked the common assumption of a constant Arrhenius prefactor in the nanosecond range, which yields dwell times over 6 orders of magnitude above our measured values.

Finally, we gave theoretical predictions of (sub)nanosecond dwell times at zero voltage under a 25~mT in-plane field, as a function of junction diameter and effective anisotropy.

By advancing the fundamental understanding of stochastic magnetization switchings in magnetic tunnel junctions, our work hereby paves the way towards the implementation of (sub)nanosecond-operating stochastic units for cognitive computing, with perpendicular magnetic stacks that can be scaled to tens of nanometers.

\begin{acknowledgments}
 We graciously thank Mark D. Stiles, Liliana Buda-Prejbeanu, Louis Hutin, Eyub Yildiz, Joo-Von Kim, Olivier Fruchart, Mair Chshiev, Helene Bea, and Olivier Boulle for fruitful discussions and suggestions. NSF-ANR supported this work via grant StochNet Project ANR-21-CE94-0002-01. This work was partially supported by NSF grant number CCF-CISE-ANR-FET-2121957, the Carnot project PRIME SPOT ANR P-22-03813, and the French RENATECH network. L. D. acknowledges funding from the University of Li{\`e}ge under Special Funds for Research, IPD-STEMA Programme.
\end{acknowledgments}

\bibliography{references_2}

\end{document}